\documentstyle[12pt]{article}

\begin{document}
\title{ Diluted random fields\\
in \\ Mixed cyanide crystals }
\author{Serge Galam}
\date{
Laboratoire des Milieux D\'{e}sordonn\'{e}s et H\'{e}t\'{e}rog\`{e}nes \footnotemark[1], 
\\ÊUniversit\'e Paris 6,
Tour 13 - Case 86, 4 place Jussieu, 75252 Paris Cedex 05, France\\
(E-mail: galam@ccr.jussieu.fr)\\$\ $\\
{\em Short title:\/} Orientational Glasses\\ {\em PA Classification Numbers:\/}
 64.60 A, 64.60 C, 64.70 P }
\maketitle
\addtocounter{footnote}{1}
\footnotetext{Laboratoire associ\'{e} au CNRS (URA n$^{\circ}$ 800)}

\begin{abstract}

A percolation argument and a dilute compressible random field Ising model are used to present 
a simple model for mixed cyanide crystals. 
The model reproduces quantitatively several features of the phase diagrams altough some
crude approximations are made.
In particular critical thresholds $x_c$ at which ferroelastic first order transitions disappear, 
are calculated. Moreover, transitions are found to remain first order down to $x_c$  for all mixtures
except for $bromine$, for which the transition becomes continuous.
All the results are in full agreement with experimental data.

\end{abstract}
\newpage

\section{Introduction}

Mixed cyanide crystals $X(CN)_{x}Y_{1-x}$ where $X$ is an alkali metal 
($K$, $Na$ or $Rb$)
and $Y$ stands for a spherical halogen ion 
($Br$, $Cl$ or $I$) exhibit puzzling phase diagrams [1].
In particular the pure alkali-cyanide $XCN$ ferroelastic transition disappears
at some cyanide concentration threshold $x_c$ which
is not equal to the associated percolation threshold $p_c$. Moreover $x_c$
varies with both $X$ and $Y$ compounds. For instance, $x_c=0.60$ and $x_c=0.80$ 
for respectively $Y=Br$ and $Y=Cl$ with $X=K$.

Below $x_c$ experimental
evidences hint to a cyanide orientational freezing, together with no
long range quadrupolar order. In parallel, the high temperature
cubic symmetry is preserved at low temperatures [1]. Analogies to spin-glasses and random 
field systems were
suggested to elucidate the orientational glass state physical nature [1, 2].

In addition, while the transition
stays first order for $X(CN)_{x}Cl_{1-x}$ [3], it was found to become
continuous at $x_c$ for $X(CN)_{x}Br_{1-x}$ [4]. These experimental facts are
still lacking an explanation.

The complicated symmetry of the molecules involved as well as the form of quadrupolar interaction
make theoretical attemps rather difficult and heavy. Previous approaches
[2 and references therein] tried to embody some of the molecule physical charateristics.
However they failed to explain above features.

In this letter a model which combines a dilute compressible random field Ising system with 
a percolation argument is suggested
to reproduce parts of the mixed cyanide phase diagrams.
A microscopic calculation of critical thresholds $x_c$ is presented.
Compressibility is found to produce
a first order transition only at cyanide concentrations larger than a threshold $x_L$. 
On the opposite, 
diluted random fields are shown to activate a first order transition solely
below another cyanide threshold $x_r$.
 
Transition orders which depend on the respective values of $x_c$, $x_L$ and $x_r$,
are discussed for a large variety of mixtures. 
We predict transitions to remain first order down to $x_c$ for all mixtures, 
except in the case of $bromine$ which exhibits a continuous transition at $x_c$.
Our results fit perfectly a series of experimental data.

\section{A percolation approach at $T=0$}

Starting from the $x_c \neq p_c$ experimental fact, we conclude that
some cyanides do not participate in the propagation of quadrupolar
long range order. These cyanides are thus ``neutralized" from reorientation.
Due to steric hindrance,
a local deformation of the unit cell is expected on
substituting one spherical $Y$ ion to the dumbbell-shaped cyanide. We assume
that a local $XY$ unit cell, embedded in $XCN$ in bulk, matchs the unit cell of $XY$ in bulk.
The associated volume deformation of the former unit cell is
$
\Delta v = \frac{1}{4}( a_{KCN}^3 - a_{KX}^3 ) \:. 
$
From symmetry, each one of the $c$ $Y$ nearest neighbors 
is affected
by this local volume deformation which thus produces the overall
volume deformation $\Delta V=c\Delta v$ ($c=12$ and $p_c=0.198$ 
on the {\sf fcc} cyanide sublattice [5]).

$\Delta V$ may be extracted from neighboring
unit cells using their 
free volumes which originate
from volume differences between the unit cell and the molecule itself [6]. 
The free volume per molecule is
$
v_f=\frac{a_{XCN}^3}{4}-(v_X+v_{CN})\:,
$
where $a_{XCN}$ is the pure $XCN$ lattice constant, 
$v_X$ and $v_{CN}$ are ion volumes [7].

Dumbbell-shaped $CN$ reorientations are directly coupled to the unit cell shape via steric 
hindrance mechanisms. 
Any volume deformation of a cyanide cage (increase or decrease) lowers the corresponding symmetry 
which in turn lowers the number of accessible orientations. Here we are
assuming that a cyanide whose cage is deformed (increased or decreased) becomes orientationally
``neutralized". 
Accordingly, one $Y$ substitution will affect on average, 
\begin{equation}
\alpha = c \frac{|\Delta v|}{v_f}\:,
\end{equation} 
unit cells by deforming their respective free volumes. 
As a consequence, we obtain an effective 
density of free to reorient cyanides,
$
x_{f}= x-\alpha(1-x) \:.
$
These free to reorient cyanides thus obey, by definition, site percolation with $x_{f,c}=p_c$ 
at $x=x_c$. Therefore we get,
\begin{equation}
x_c = \frac{p_c+\alpha}{1+\alpha}\:.
\end{equation}

Without a fitting parameter the calculation
of $x_c$ is readily performed using crystallographic data [7, 8, 9].
The results are obtained for various mixtures (see the Table) including systems for which no 
experimental data are available.
Below $x_c$ a region of randomly oriented
ferroelastic domains with no static phase transition is predicted.
These domains will shrink with increasing dilution to disappear eventually when $x_{f}=0$
at a new threshold [10],
$
x_{d}=\frac{\alpha}{1+\alpha}\:.
$
Shear torque experiments [11], as well as diffraction experiments [12] suggested a very 
similar phase diagram.
\section{A crude Ising-like model}

To extend above results to $T\neq 0$ we now build the simplest possible model Hamiltonian
making several crude approximations.
\begin{itemize}

\item From symmetry, cyanides have several equivalent orientations making
$q$-Potts variables appropriate 
[13]. However we restrict ourselves to the minimum number of orientations 
required to sustain
an orientational long range order, i.e.,
two. Therefore we use Ising variables 
$\{S_i=\pm 1\}$ to mimic the cyanide orientational degrees of freedom.

\item Cyanide quadrupolar interactions are long ranged. However
our percolation argument is a short range effect. Therefore, to be consistent with above $T=0$
calculation, we consider short range ferromagnetic couplings.

\item To fit the first order character of the pure $XCN$ transition, we introduce elastic 
degrees of freedom in order to turn the
continuous Ising ferromagnetic transition to first order. To keep calculations simple,
we use an harmonic model of volume fluctuations though 
it is clearly a rudimentary model of elasticity [14]. 

\item Within a model of two equivalent orientations, a cyanide prevented from reorientation 
is trapped along one direction. To embody this effect, we introduce local quenched
random fields.
The probability $p_{t}$ to have a local random field is equal to the probability of 
having a deformed cyanide cage.
From the density $x_t$ of trapped cyanide we obtain,
\begin{equation}
p_{t}= \left\{ \begin{array}{ll}
                 \frac{\alpha(1-x)}{x} & \mbox{if}\: x_{d} \leq x \leq 1 \\
                  1                    & \mbox{if}\: x < x_{d}
                 \end{array}
       \right. \:.
\end{equation}
The distribution function for the random filed $h_{i}$ is then,
\begin{equation}
P(h_{i})=\frac{p_{t}}{2}[\delta (h_{i}-h)+\delta (h_{i}+h)]+(1-p_{t})\delta(h_{i})\:. 
\end{equation}
It satisfies both required symmetry conditions $P(h_{i})=P(-h_{i})$ and $\overline{ h_{i}}=0$  
(the overline denotes a field configurational average) which preserve the cubic symmetry of 
the associated problem. 

\end{itemize}

\section{The Hamiltonian}

From above approximations, we obtain the following 
effective Hamiltonian,
 \begin{equation}
H_{eff} = -G \sum_{<i,j>}\epsilon_{i}\epsilon_{j}S_iS_j
-\frac{E}{N^d} (\sum_{<i,j>}
\epsilon_{i}\epsilon_{j}S_iS_j)^2 -\sum_{i}\epsilon_{i}h_{i}S_{i}\:,
\end{equation}
where $G$ and $E$ are constants, $N$ is the total number of spins, $d$ is the dimension [15], 
and $\epsilon_{i}$ is a random site
variable. It is $1$ if site $i$ is occupied by a cyanide and $0$ otherwise. 
We have $\{\epsilon_{i}\}_{av} = x$, where $\{...\}_{av}$ denotes a configurational average 
over site disorder. 

We define the order parameter as $m=\{\overline{<\epsilon_{i}S_{i}>}\}_{av}$ where averages are 
taken over both site and 
random field disorders.  
The associated mean field site free energy is,
\begin{eqnarray}
{\cal F}&=&\frac{1}{2} cGm^{2} + \frac{3L}{4}m^4 \nonumber \\
        & &  -x k_{B}T \left [ \frac{p_{t}}{2}\{ln[cosh(\beta cGm
+\beta Lm^{3}+\beta h)] +
ln[cosh(\beta cGm+\beta Lm^{3}-\beta h)]\} \right. \nonumber \\
        & &  +(1-p_{t})ln[cosh(\beta cGm+\beta Lm^{3}){\bf ]}\left ]
\,\frac{\,\,\,}{\,\,\,} \,
 k_{B}Tln(2) \right.\:,
\end{eqnarray}
where $\beta\equiv\frac{1}{k_BT}$, $k_B$ is the Boltzman constant, $T$ is the temperature and
$L=c^2E$ (more details will be published elsewhere).
In real systems $h_{i}$ and $\epsilon_{i}$ are correlated. However, in order to be consistent 
with the previous short range interaction approximation, to keep the calculations simple
and doing a mean field calculation, 
these correlations are neglected here.
We now analyse two simple limiting cases which are physically meaningfull.
\subsection{The zero-steric hindrance effect case ($\alpha=0$)}
From a Landau expansion of Eq. (6) a continuous transition
occurs at the critical temperature, $k_{B}T_{c}=xcG$
under the condition of a positive quartic coefficient,
$
B=-\frac{L}{cG}+\frac{1}{3x^2}\:,
$
which results in the condition $x<  x_L$
on cyanide concentration [16] where,
\begin{equation}
x_L\equiv  (\frac{cG}{3L})^{1/2}\:.
\end{equation}
At $x=x_L$ the transition turns first order via a 
tricritical point ($B=0$ with a positive free energy sixth order coefficient). 

At $x=1$, $p_t=0$ (Eq. (3)) even if $\alpha \neq 0$. From experimental works pure $XCN$ exhibits 
a first order transition. 
All plastic systems must therefore satisfy  $x_L < 1$ with thus $L>\frac{cG}{3}$.
Dilution weakens
the first order character of the transition.
The associated negative quartic term $B$ becomes smaller in amplitude to vanish eventually at a 
tricritical point ($x=x_L$). 
There, the transition is continuous with tricritical exponents. Upon further dilution the 
transition becomes second-order with $0\leq x< x_L$.

\subsection{The zero-compressibility case ($L= 0$)}
In this case, random field are acting alone ($\alphaÊ\neq 0$). 
At $x=1$, with $p_t$ being an independent 
external parameter, the zero-compressibility free energy becomes identical to that of the trimodal 
random field Ising model [17]. 
A first order transition is found only for
$0.73<p_t\leq 1$ and for some restricted range of random field intensities 
$\sim 0.55 < \frac{h}{cG}< \sim 0.65$ [17].
These results produce an additional threshold in cyanide density around $p_t=0.73$ (Eq. (3)),
\begin{equation}
x_r\equiv \frac{\alpha}{0.73+\alpha}\:.
\end{equation} 
Only at $x<x_r$ can dilution turn
the transition to first order via random fields. However condition $x_c<x_r$ must also be satisfied
since the transition itself disappears at $x_c$.
The equivalent constraint on $\alpha$ gives,
\begin{equation}
\alpha >\frac{0.73 p_c}{1-0.73-p_c}\sim 2.09\:.
\end{equation} 

 Eq. (8) gives $x_r=0.60$ and $x_r=0.83$ for respectively $K(CN)_{x}Br_{1-x}$ ($\alpha=1.12$)
 and $K(CN)_{x}Cl_{1-x}$
($\alpha=3.48$). In parallel the Table shows $x_c=0.62$ and $x_c=0.82$ 
for $Br$ and $Cl$ respectively.
On this basis we conclude that upon dilution, while random fields can turn the
transition first order in $K(CN)_{x}Cl_{1-x}$ mixtures ($x_c<x_r$), they cannot 
do it for $K(CN)_{x}Br_{1-x}$ mixtures
($x_c>x_r$).

\section{Conclusion}

In our model, both compressibility and random field drive
the transition to first order. However, the compressibility effect weakens with dilution to fade out in 
the vicinity of $x_c$ for all systems. 
On the opposite, random fields start to be active
at dilution below $x_r$, which is possible only when $x_r>x_c$. Using the condition $\alpha >2.09$
(Eq. (9)), the Table shows it is always satisfied except for mixtures with $bromine$.

We can thus predict that dilution with $chlorine$ and $iodine$ maintains
the transition first order down to $x_c$. Only $bromine$ turns the transition continuous.
A tricritical point is expected for $X(CN)_{x}Br_{1-x}$ mixtures.
Our predictions reproduce
experimental results with respect to $potassium$ systems [3, 4]. 
Additional experiments on mixtures with
$sodium$ and $rubidium$ would provide a definite ground to our model. 

It was worth to stress that the extension of the model to $q$-Potts variables 
will not discard  elasticity with respect to the first order character of the transition.
Indeed, the rigid Potts model does exhibit a first order 
transition at $d=3$ , but only when $q> 2$. It is the case for $KCN$, but not for
all systems. In particular, most materials which have a tetrahedral molecule in a cubic site
will have equivalent orientations with $T_d$ symmetry, i.e., $q=2$, like for instance
the perchlorate tetrahedron in $KClO_4$. Therefore, elasticity is a necessary feature of our
model.

Last but not least, the predictive power of our model comes as 
a surprise according to the several approximations made. In particular the short range character 
of the interactions is hard to justify. However,
the results are impressive showing there must exist some screening mechanism 
which allows such a simple model.

\subsection*{Acknowledgments}
I would like to thank Ph. Depondt, P. Doussineau and A. Levelut
for stimulating comments on the manuscript.

\newpage

\begin{table}
\caption{\sf Numerical values calculated for $\alpha$, $x_c$ and $x_d$ and experimental 
thresholds when known (denoted by `{\it exp:}'). Units for lengths
and volumes
are $\AA$  and $\AA^3$.
See details in the text. Error bars are within respectively $\pm 0.01$ for all data in the Table,
and $\pm 0.05$ 
for experimental thresholds.} 
\label{tbl}
\begin{tabular}{|l|l|l|l|l|l|l|l|} \hline  $XCN/ Y$ &$a_{XCN}$ & 
$a_{XY}$ &$\Delta v$ &$v_f$ &$\alpha$&$x_c$&$x_d$\\  \hline 
$KCN/ Cl$ & 6.53 &6.29 & 7.30& 25.19& 3.48& 0.82 {\it exp:} 0.80& 0.78 
{\it exp:} 0.75 \\ 
$KCN/ Br$ & 6.57 &6.60 & 2.36& 25.19& 1.12& 0.62 {\it exp:} 0.60& 0.53 
{\it exp:} 0.50 \\ 
$KCN/ KI$ & 6.53 &7.06 &18.46& 25.19& 8.79& 0.92 {\it exp:} 0.90& 0.90\\  
$NaCN/ Cl$& 5.90 &5.65 & 5.99& 12.67& 5.68& 0.88	{\it exp:}$\sim 0.80$ & 0.86 \\ 
$NaCN/ Br$& 5.90 &5.97 & 2.11& 12.67& 1.99& 0.73	& 0.67 \\    
$NaCN/ KI$ & 5.90 &6.47 &16.63& 12.67& 15.74& 0.95	& 0.94 \\ 
$RbCN/ Cl$& 6.82 &6.58 & 8.05& 30.98& 3.12& 0.80	& 0.76 \\ 
$RbCN/ Br$& 6.82 &6.85 & 1.19& 30.98& 0.46& 0.45 {\it exp:} 0.55 & 0.32 \\ [5pt] \hline
\end{tabular}
\end{table}

\end{document}